\documentclass[twocolumn,prb,showpacs,multicol,amsmath,amssymb]{revtex4}
\usepackage[dvips]{graphicx}
\usepackage{graphicx}
\usepackage{dcolumn}
\usepackage{bm}
\usepackage{graphics}
\usepackage{epsfig,color}

\newcommand{\be}{\begin{equation}}
\newcommand{\ee}{\end{equation}}
\newcommand{\bea}{\begin{eqnarray}}
\newcommand{\eea}{\end{eqnarray}}

\newcommand{\bS}{\bf S}

\begin{document}

\title{Thermodynamics of the spin-1/2 two-leg ladder compound $(C_{5}H_{12}N)_{2} CuBr_{4}$}

\author{Fatemeh Amiri$^1$, Saeed Mahdavifar$^{1\star}$, Mahboobeh Shahri Naseri$^2$}
\affiliation{ $^1$Department of Physics, University of
Guilan, 41335-1914, Rasht, Iran\\
$^2$Department of Physics, Payam Noor University
, 19395-3697, Tehran, Iran}
\email[]{mahdavifar@guilan.ac.ir, smahdavifar@gmail.com}
\date{\today}

\begin{abstract}

 The thermodynamic behavior of the spin $S=1/2$  antiferromagnetic two-leg  ladder compound $(C_{5}H_{12}N)_{2} CuBr_{4}$ in a uniform magnetic field is studied using numerical and analytical approaches.  The entropy $S(H,T)$ and specific heat $C(H,T)$ are calculated. The specific heat shows various behaviors in different regions of the magnetic field. The field-dependence of the specific heat is almost symmetric about the average of quantum critical fields in complete agreement with experimental results. In addition, it is found that during an adiabatic demagnetization process, temperature drops in the vicinity of the field induced zero-temperature quantum phase transitions.

\end{abstract}

\pacs{75.10.Jm; 75.10.Pq}

\maketitle


\section{Introduction}\label{sec1}

Currently, wide interest is devoted to the low-dimensional gapped spin systems, both experimentally and theoretically. In particular, magnetic spin ladders\cite{Dagotto96} have attracted enormous attention in recent years, since remarkable progress in the fabrication of such ladder compounds. Magnetic spin ladders which intermediate between one-dimensional and two dimensional spin systems, have a gapped or gapless ground state, respectively, for an even or an odd number of legs\cite{Dagotto96, Dagotto93}. The two-leg spin ladder Hamiltonian in the presence of a magnetic field is defined as

\begin{eqnarray}
{H} &=& J_{\parallel} \sum_{n,\alpha} {\bS}_{n,\alpha} \cdot
{\bS}_{n+1,\alpha} - g \mu_{B} B  \sum_{n,\alpha}
S^{z}_{n,\alpha} \nonumber\\
&+& J_{\perp} \sum_{n}{\bS}_{n,1} \cdot {\bS}_{n,2} \,  \label{Hamiltonian}
\end{eqnarray}
where $\bS_{n,\alpha}$ is the spin $S=1/2$ operator on rung $n$
($n=1,...,N/2$) and leg $\alpha$ ($\alpha=1,2$). An applied field $B$ in the $z$ direction leads to Zeeman term. There is a large body of research on the so-called telephone number compounds like $Sr_{14}Cu_{2}O_{41}$ which are also cuprate ladders like
$SrCu_{2}O_{3}$. Several classes of materials such  $(5IAP)_2CuBr_4.2H_2O$, $Cu_{2}(C_{5}H_{12}N_{2})_{2}Cl_{4}$ and
$(C_{5}D_{12}N)_{2}CuBr_{4}$ can be also well described experimentally by the two-leg Heisenberg antiferromagnetic ladder model\cite{Azuma94, Landee01, Chaboussant97-1, Chaboussant98-8, Chaboussant97-2, Savici09}. Specially, piperidinium copper bromide $(C_5H_{12}N)_{2}CuBr_{4}$ is known as one of the best spin-1/2 ladder compounds with large rung exhange\cite{Patyal90,Watson01,Lorenz08,Klanjsek08,Anfuso08,Ruegg08,Thielemann09-1,Thielemann09-2,Sologubenko09, Cizmar10}. It is found that the rung exchange $J_{\perp}=12.9(2) K$ is about four times larger than  $J_{\parallel}=3.3(2) K$. In the absence of the magnetic field, the excitation spin gap is equal to $\simeq 9.5 K$. In the presence of a magnetic field, the system remains in a gapped disordered phase below $B_{c1}\simeq6.99(5)~T$. The spin excitations are gapless and the system is in the Luttinger-liquid (LL) phase in the intermediate region, $B_{c1}<B<B_{c2}$. At the second critical field, $B_{c2}\simeq14.4(1)~T$, system undergoes a phase transition to the fully polarized state. Also, the LL state extends down to the temperature of a 3D magnetic ordering transition at $T_{N}\leq 110~mK$, suggesting an intermediate coupling $J'\simeq27 mK\ll J_{\perp},J_{\parallel}$. The LL predictions have been quantitatively tested for magnetization and specific heat\cite{Ruegg08}, nuclear magnetic reasonance\cite{Klanjsek08}, neutron differaction\cite{Thielemann09-1} and thermal conductivity\cite{Sologubenko09} measurements.

There are many research works dedicate to the zero-field properties of the ladder system, while current interest has been focused on the temperature dependent behaviors and  especially the magnetocaloric effects (MCE) in the spin ladder which has not been studied completely so far. It is believed that the low temperature behaviors give important insight in the physics of quantum phase transitions. In this paper, we use both theoretical and numerical techniques to prepare a complete picture of the dependence of entropy, specific heat and adiabatic (de)magnetization on both magnetic field and temperature. Our numerical approach is based on a full-diagonalization method while we use Jordan-Wigner transformation analysis for the analytical part. We show that our results agree very well with the experiment.

The outline of the paper is as follows. In section II we discuss the model in the strong antiferromagnetic coupling
limit and derive the effective spin chain Hamiltonian. The Jordan-Wigner transformation for the spin $S=1/2$ effective \emph{XXZ} chain Hamiltonian will be considered and the basic mean-field set up will be presented in section III. In section IV, we present results of the mean-field and full diagonalization calculations and compare with the experimental data. Finally, we conclude and summarize our results in section V.


\section{Effective Model} \label{sec2}

The thermodynamics of the system can be analyzed using a mapping of the spin $S=1/2$ ladder onto the effective spin chain in the strong-coupling limit\cite{Mila98, Totsuka98, Chaboussant98}. The first-order degenerate perturbation theory in the parameter $J_\perp$ and $J_\parallel$, must be done to derive an effective low-energy model. Since, the whole magnetization curve has a width which is related to the bandwidth of the triplet excitation, the first-order perturbation theory is sufficient. At $J_\perp \gg J_\parallel$, it is convenient to discuss the model by representing the site-spin algebra in terms of on-bond-spin operators. In the limit, $J_\parallel=0$ the ladder divided into the isolated rungs. Indeed an isolated rung may be in a
singlet or a triplet state with corresponding spectrum given by
\begin{eqnarray}
E_{\pm}=(\frac{J_{\perp}}{4}\pm g \mu_{B} B),~~   E_{0}=\frac{J_{\perp}}{4},~~   E_{s}=\frac{-3J_{\perp}}{4}.
\end{eqnarray}
By increasing the magnetic field, one component of the triplet becomes closer to the singlet ground state and in a strong enough magnetic field, $g\mu_{B} B=J_{\perp}$, the triplet $|\uparrow\uparrow\rangle$ is exactly degenerate with the singlet. In principle, we have a situation that the singlet and the triplet, $|\uparrow\uparrow\rangle$, create a new effective spin $S=1/2$ system. One can easily project the original ladder Hamiltonian (1) on the new singlet-triplet subspace
\begin{eqnarray}
|\Uparrow\rangle&=&\frac{1}{\sqrt{2}}(|\uparrow\downarrow\rangle-|\downarrow\uparrow\rangle),\nonumber \\
|\Downarrow\rangle&=&|\downarrow\downarrow\rangle.
\end{eqnarray}
This leads to the definition of the effective spin $S=1/2$ operators
\begin{eqnarray}
S^{\dag}_{n,\alpha}  &=& (-1)^{n+\alpha} \frac{1}{\sqrt{2}}\tau^{\dag}_{n}, \nonumber\\
S^{z}_{n,\alpha}  &=& \frac{1}{4}(\emph{I}+2\tau^{z}_{n}).
\end{eqnarray}
The effective Hamiltonian in terms of the effective spin operators up to the accuracy
of an irrelevant constant becomes the Hamiltonian of the spin $S=1/2$  anisotropic \emph{XXZ} chain in an effective magnetic field
\begin{eqnarray}
H^{eff} &=& J_{\parallel} \sum_{n} (\tau^{x}_{n}.\tau^{x}_{n+1}+ \tau^{y}_{n}. \tau^{y}_{n+1}+ \Delta \tau^{z}_{n}. \tau^{z}_{n+1}) \nonumber\\
&-& g \mu_{B} B^{eff} \sum_{n} \tau^{z}_{n}.
\end{eqnarray}
which allows for rigorous analysis. The $\Delta=1/2$ is the anisotropy parameter and $B^{eff}=B-\frac{2 J_{\perp}+J_{\parallel}}{2 g \mu_{B}}$. At zero temperature, the gapped disordered phase in the ladder system corresponds to the negatively saturated magnetization phase for the effective
spin chain, whereas the LL phase of the main ladder system corresponds to the finite magnetization
phase of the effective spin-1/2 chain. The second critical field where the ladder system is totally
magnetized, corresponds to the fully magnetized phase of the effective spin chain.


\section{Fermionization} \label{sec3}

Theoretically, the energy spectrum is needed to investigate the thermodynamic properties of the model. In this respect, we implement the Jordan-Wigner transformation to fermionize
the effective $XXZ$ model. Using the Jordan-Wigner transformation

\begin{eqnarray}
S^{+}_{n} &=& e^{i\pi\sum ^{n-1} _{m} c^{\dag}_{m} c_{m}}  c^{\dag}_{n}, \nonumber\\
S^{z}_{n} &=& c^{\dag}_{n} c_{n} - \frac{1}{2}.
\end{eqnarray}
 the effective Hamiltonian is mapped onto a 1D model of interacting spinless fermions
\begin{eqnarray}
H_{f}^{eff}&=& J_{\parallel} \sum_{n} [\frac{1}{2}  (c^{\dag}_{n}c_{n+1}+c^{\dag}_{n+1}c_{n})+ \Delta  c^{\dag}_{n}c_{n}c^{\dag}_{n+1}c_{n+1}]\nonumber \\
&-& (J_{\parallel}\Delta+g \mu_{B} B^{eff}) \sum_{n} c^{\dag}_{n}c_{n}\nonumber\\
&+& \frac{N}{4}(J_{\parallel}\Delta+2 g \mu_{B} B^{eff}). \nonumber\\
\end{eqnarray}
This Hamiltonian is not exactly solvable because the fermion interaction. Therefore the mean-field theory is applied to interaction term\cite{Dmitriev02}. The fermion interaction term is decomposed by mean-field parameters which are related to spin-spin correlation functions as
\begin{eqnarray}
\gamma_{1}&=&<c^{\dag}_{n}c_{n}>,\nonumber \\
\gamma_{2} &=& <c^{\dag}_{n}c_{n+1}>,\nonumber \\
\gamma_{3} &=& <c^{\dag}_{n}c^{\dag}_{n+1}>.
\end{eqnarray}
Utilizing the above order parameters and perform a Fourier transformation to momentum space by using $c_{n} = \frac{1}{\sqrt{N}} \sum ^{N} _{n=1} e^{-ikn} c_{k}$, the mean field Hamiltonian is given by:

\begin{eqnarray}
H_{MF}^{eff} &=&  \varepsilon_{0}+\sum_{k>0} a(k) (c^{\dag}_{k}c_{k}+c^{\dag}_{-k}c_{-k})\nonumber \\
&-& i  \sum_{k>0} b(k) (c_{k}c_{-k}+c^{\dag}_{k}c^{\dag}_{-k}),
\end{eqnarray}
where,
\begin{eqnarray}
\varepsilon_0&=&N(\frac{1}{2} g \mu_{B} B^{eff}+\Delta J_{\parallel} (1/4+\gamma_{1}^{2}+\gamma_{3}^{2}-\gamma_{2}^{2})),\nonumber \\
a(k)&=&J_{\parallel}(1-2 \gamma_{2} \Delta)\cos(k)+J_{\parallel}\Delta(2 \gamma_{1}-1)\nonumber \\
&-& g \mu_{B} B^{eff}, \nonumber \\
b(k)&=&2 J_{\parallel} \gamma_{3} \Delta \sin (k).
\end{eqnarray}
Using the following unitary transformation
\begin{eqnarray}
c_{k}=\cos(k) \beta_{k}- i\sin(k) \beta^{\dag}_{-k},
\end{eqnarray}
the diagonalized Hamiltonian is given by
\begin{eqnarray}
H_{MF}^{eff} &=& \varepsilon_{0} +\sum_{k} \varepsilon(k) (\beta^{\dag}_{k} \beta_{k}-\frac{1}{2}),
\end{eqnarray}
where $\varepsilon(k)$ is the dispersion relation
\begin{eqnarray}
\varepsilon(k) &=&  \sqrt{a^{2}(k) + b^{2}(k)}.
\end{eqnarray}
In order to solve mean-field Hamiltonian, the following self-consistent equations should be satisfied
\begin{eqnarray}
\gamma_{1} &=&\frac{1}{2\pi} \int^{\pi}_{-\pi}[\frac{a}{\varepsilon}(\frac{1}{1+\exp(\beta \varepsilon)})+\frac{1}{2}(1-\frac{a}{\varepsilon})]dk, \nonumber\\
\gamma_{2} &=& \frac{1}{2\pi} \int^{\pi}_{-\pi}\cos(k)[\frac{a}{\varepsilon}(\frac{1}{1+\exp(\beta \varepsilon)})+\frac{1}{2}(1-\frac{a}{\varepsilon})]dk, \nonumber\\
\gamma_{3} &=&\frac{1}{2\pi} \int^{\pi}_{-\pi} \sin(k)\frac{b}{\varepsilon}(\frac{1}{1+\exp(\beta \varepsilon)}-\frac{1}{2}) dk.
\end{eqnarray}
Using the above order parameters, the thermodynamic
functions such as the free energy, entropy and the specific heat
are expressed as
\begin{eqnarray}
f &=& \varepsilon_{0}-\frac{1}{2 \pi K_{B}T} \int^{\pi}_{0} dk \ln (2 \cosh (\frac{\varepsilon(k)}{2K_{B}T})),\nonumber \\
S&=&\beta^{2}\frac{\partial f}{\partial \beta}, \nonumber \\
C&=&-K_{B} \beta \frac{\partial S}{\partial \beta}.
\end{eqnarray}

\section{Results} \label{sec4}

In this section we present our results obtained by analytical fermionization approach and the numerical full diagonalization technique on small size systems ($N=8, 12, 16$). In the following, using both numerical and analytical methods the field and temperature dependent properties of the compound piperidinium copper bromide $(C_{5}H_{12}N)_{2} CuBr_{4}$ are studied.

\subsection{Specific heat}

\begin{figure}[t]
\centerline{\psfig{file=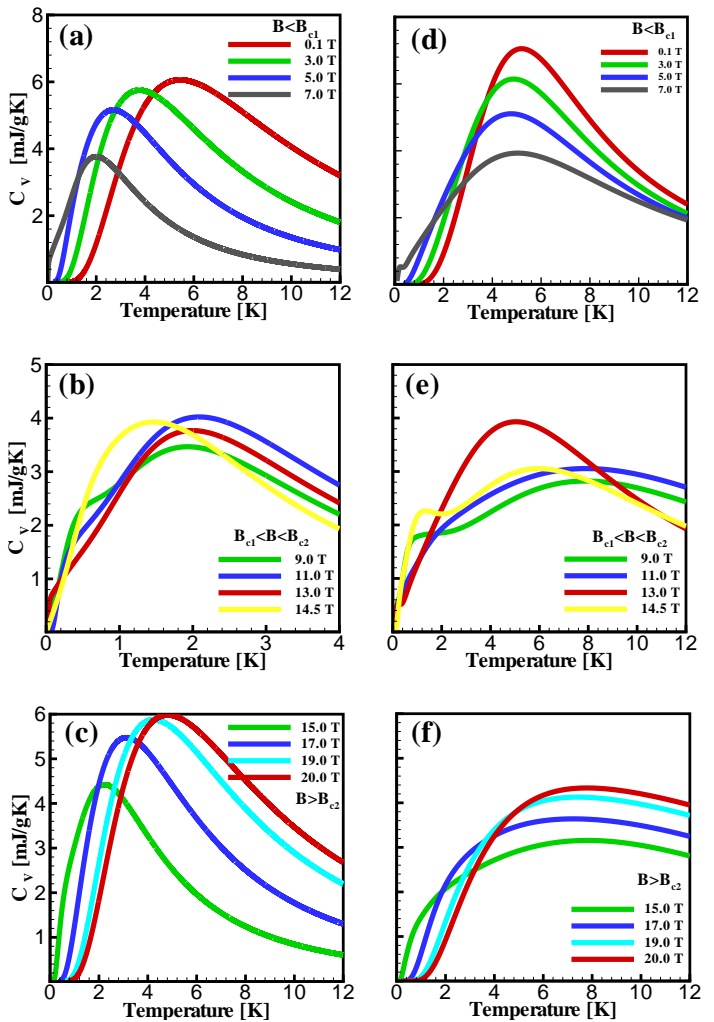,width=3.65in}}
\caption{(color online). The specific heat as a function of temperature in the fixed magnetic field for ladder with
$J_{\perp}=13.0 K$, $J_{\parallel}=3.25 K$.\\
(a-c)mean-field ~~~~~ (d-f)full diagonalization for $N=16$.}
\label{1}
\end{figure}

In Fig.~\ref{1}, we focus on the behavior of the temperature dependence of the specific heat for $(C_{5}H_{12}N)_{2} CuBr_{4}$. It shows three regimes: quantum disordered(QD), spin LL and fully polarized phase. Fig.~\ref{1} (a)-(c) and (d)-(f) shows the mean-field and full diagonalization results of specific heat, respectively. As shown in Fig. \ref{1} (a) and (d), in QD regime,  $B<B_{c1}$, the specific heat decays exponentially due to the presence of spin gap. It shows a single peak and when field increases the peak shifts to lower temperature with the height decreased.
This peak is related to the triplet excitations of the ladder.
As the field B reaches to $B_{c1}$, the spin gap is reduced and a shoulder gradually emerges at low temperature, which is a signature of approaching the quantum critical point\cite{sachdev99, sachdev96}.

When $B_{c1}<B<B_{c2}$ (Fig.~\ref{1}(b) and (e)), as $B$ is increased further, a new peak becomes visible that in the middle of the LL, we can clearly see two peaks in the temperature dependence of the specific heat. Below the first peak, the temperature dependence remains linear up to $B_{c2}$. The linearity of C is shown in Fig.~\ref{1}(b) and (e). By increasing the field, height of the first peak is decreased but its position is almost unchanged, while the second peak shifts to the high-temperature as illustrated in Fig.~\ref{1}(e) and yielding a very good agreement with the experimental results\cite{Ruegg08}.
When $B$ approaches $B_{c2}$, the peak in low temperature starts to vanish and just a shoulder can be seen (Fig.~\ref{1}(b) and (e)).
\begin{figure}[t]
\centerline{\psfig{file=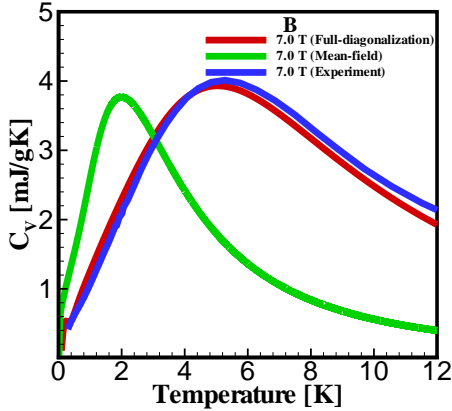,width=2.65in}}
\caption{(color online). Comparison of the specific heat as a function of temperature in fixed magnetic field with experiment (Ref.[\onlinecite{Ruegg08}]) for ladder with
$J_{\perp}=13.0 K$, $J_{\parallel}=3.25 K$.}
\label{2}
\end{figure}
For $B>B_{c2}$, fully polarized state, the specific heat decays exponentially as $T\rightarrow0$   because of opening of the gap. Near the critical field $B_{c2}$, the shoulder that emerges in the LL vanishes and the specific heat has a single peak. By more increasing the field, $B>B_{c2}$, the gap and the height of the peak increases and the position of peak shifts slightly to higher temperatures. The result shows that the system is in a few different states under different magnetic fields (Fig.~\ref{1}(c) and (f)).

We should note that, this model is experimentally studied recently\cite{Ruegg08}. The calculated values of the specific heat with both techniques are compared with the experimental results. The results from full-diagonalisation have a quantitative accuracy within $2-8\% $ compared to the experiment. At low temperature region, our results agree very well with the experimental data. But at higher temperatures, the quantitative agreement of results due to the mean-field approach is missing, as shown in Fig.~\ref{2}. It is mostly because of considering the effective XXZ Hamiltonian and missing some of the high energy spectrum of the ladder system.

The excitation is a triplet and the effective XXZ chain model considers only one of its three branches. Thus this model is really intended only for the regime of massless excitations, at significant fields. It works also at high fields, but certainly should not be applied at zero field. There are also two-triplet excitations, which will be missed if the one-triplet excitations are missing, but these contribute less.
In addition, the mean field approximation is used to diagonalize the effective Hamiltonian.

Now let us look at the field dependence of the specific heat. The Fig.~\ref{2}  shows the field dependence results of the specific heat obtained by mean-field approach (the same as Fig. (3)(a) of Ref.[\onlinecite{Sologubenko09}]). For temperature $T<1 K$, we can see that for $B<B_{c1}$ and $B>B_{c2}$, the specific heat enhances when temperature increases. In the LL regime, the specific heat decreases by increasing of the temperature which shows the inverse behavior as compared to low temperature. Also in this region, we observe an interesting behavior. The field dependence of specific heat is almost symmetric about $B=\frac{B_{c1}+B_{c2}}{2}=10.7 T$\cite{Ruegg08}. In the strong-coupling limit, $\frac{J\perp}{J\parallel}\gg1$, perfect symmetry would be expected due to the exact particle-hole symmetry to the XXZ chain in the magnetic field\cite{chaboussant98, Tutsuka98}.

\begin{figure}[t]
\centerline{\psfig{file=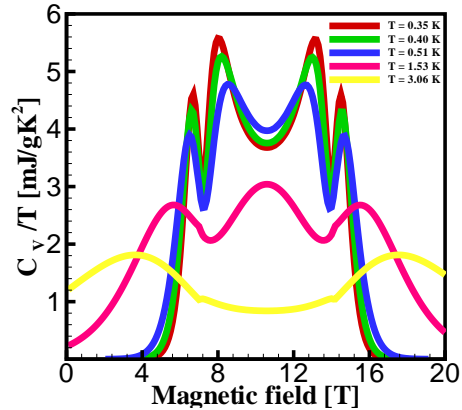,width=2.65in}}
\caption{(color online). The specific heat as a function of magnetic field in fixed temperature, for ladder with
$J_{\perp}=13.0 K$, $J_{\parallel}=3.25 K$ (the same as Fig. (2)(a) of Ref.[\onlinecite{Sologubenko09}]).}
\label{3}
\end{figure}

By increasing the temperature, the peaks starts to vanish until in high temperatures, the specific heat behaves as an ascending function, as illustrated in Fig.~\ref{3}.
\begin{figure}[t]
\centerline{\psfig{file=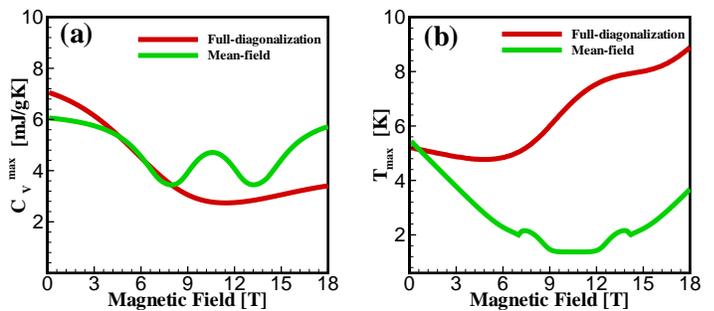,width=3.65in}}
\caption{(color online). The maximum (a) specific heat, $C_{V}^{max}$ and (b) the corresponding temperature $T_{max}$
 versus magnetic field for ladder with
$J_{\perp}=13.0 K$, $J_{\parallel}=3.25 K$.}
\label{4}
\end{figure}

It is also interesting to see the magnetic field effects on the maximum specific heat $C_{V}^{max}$ and corresponding temperature $T_{max}$. As shown in the Fig.~\ref{4}(a), $C_{V}^{max}$ obtained by both analytical and numerical approaches first decreases which shows a slow response to the splitting with applying to the magnetic field. For the magnetic fields more than $B_{c1}$, the analytical results are completely different from the numerical results. At $B=B_{c1}$ and $B=B_{c2}$, $C_{V}^{max}$ arrives at a minimum and also in the middle of the LL phase, $C_{V}^{max}$ shows a maximum at $B=\frac{B_{c_{1}}+B_{c_{2}}}{2}=10.7 T$ in our analytical results. But the numerical results show $C_{V}^{max}$ arrives at a minimum at $B=\frac{B_{c1}+B_{c2}}{2}=10.7 T$ and in good agreement with the numerical TMRG results\cite{wang00}.
Moreover, we found that the curvature of our numerical results on $T_{max}$ changes its sign at the critical fields. But again, we see that the behavior of the analytical results is completely different from our full diagonalization findings, since our analytical results are obtained for the effective XXZ chain Hamiltonian which is only valid for very low temperatures as we mentioned above.

\begin{figure}[t]
\centerline{\psfig{file=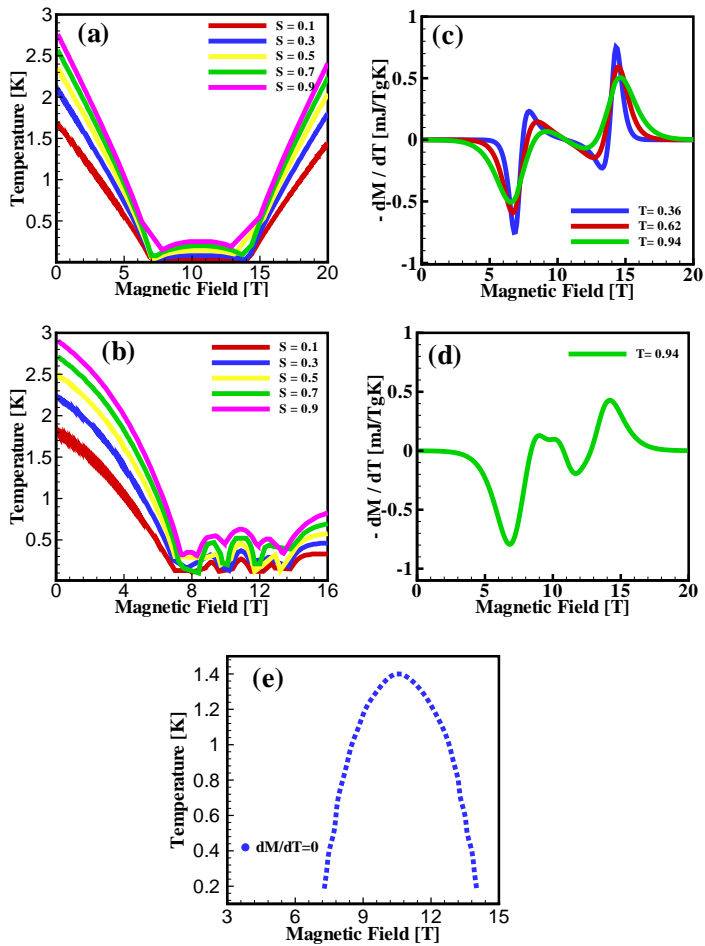,width=3.65in}}
\caption{(color online). Magnetocaloric effect for ladder with
$J_{\perp}=13.0 K$, $J_{\parallel}=3.25 K$. (a-b) Isentropes, i.e., adiabatic demagnetization curves of the $S = 1/2$ two-leg ladder. The corresponding values of the entropy are S = 0.1, 0.3, 0.5, 0.7 and 0.9 (bottom to top). (c-d) The derivative of the magnetization with respect to temperature as a function of magnetic field in fixed temperature. (e) $(\partial M/\partial T)=0$ as a function of magnetic field and temperature obtained by mean-field approximation. (a-c mean-field), (b-d full-diagonalization for $N=16$).}
\label{5}
\end{figure}

\subsection{The Magnetocaloric effect}

When a crystal containing ions is placed in a magnetic field the adiabatic or isentropic change of this external parameter causes a temperature change in the sample. This is called the magnetocaloric effect (MCE) which was first observed by Warburg\cite{warburg81}. The MCE in spin systems has attracted enormous attention in recent years. It is nowadays interesting in several aspects. From one hand, field-induced quantum phase transitions lead to universal responses when the applied field is changed adiabatically\cite{zhu03, Garst05, Honecker09, zhitomirsky04}. On the other hand, it was observed that MCE is increased by geometric frustration\cite{stat04, wessel06, schmidt07, pereira09}, promising improved efficiency in low temperature cooling applications\cite{Tsokol05, Tishin03}. More generally, the MCE is particularly large in the vicinity of quantum critical points. Here, the special focus is on the magnetocaloric effect. It will show that in this model, two minimum in the isentropes on the zero-temperature quantum critical fields can be observed. Further information about the low temperature behavior of substances can be obtained by calculating the entropy.
In this section the entropy of spin $1/2$ two-leg ladder in more detail will be discussed.

First, the mean-field and full-diagonalization results for the value of the entropy as a function of $B$ and $T$ is shown in Fig.~\ref{5}(a) and (b), respectively. The curves have been obtained by calculating the entropy, $S$, in the $B-T$ plane and determining the constant entropy curves (the isentropes) from this data.

  Fig.~\ref{5} (a) and (b) can be divided into the gapfull and gapless regimes. In the gapfull regimes,$B<B_{c1}$ and $B>B_{c2}$, a significant temperature changes happens. Firstly, for $B<B_{c1}$, by raising the magnetic field from $B=0$ to $B=B_{c1}$, the temperature can be minimized in the vicinity of $B_{c1}$. Secondly, for $B>B_{c2}$, by decreasing the magnetic field from high magnetic field to the saturation field adiabatically (adiabatic demagnetization), the temperature arrives at minimum close to the saturation field. For example, an adiabatic process which starts at $(B, T)\approx (5.5T, 0.44 K)$ or $(15.5T, 0.4 K)$ would go down to $T \approx 10 mK$  as $B\rightarrow 7.3 T$ or $B\rightarrow 13.9 T$, respectively. This case corresponds to the entropy $S=0.1$, the lowest curve in the Fig.~\ref{5} (a) and (b). For the larger value of $S=0.3$ an adiabatic process would still cool to a minimum temperature $T \approx 30 mK$. In the regime $B_{c1}<B<B_{c2}$, the energy spectrum is gapless. Therefore, small temperature changes induced by adiabatic (de)magnetization is observed. This indicates that one can use the gap-closing at the field-induced quantum phase transitions for cooling down samples.

  One can also calculate the derivative of the magnetization with respect to temperature by using the relation $(\frac{\delta Q}{\delta B})|_{T} / T=-(\frac{\partial M}{\partial T})|_{B}$. In this relation $\delta Q$ is the amount of heat which is created or absorbed by the sample for a field change $\delta B$ due to MCE. As shown in the Fig.~\ref{5}(c) and (d), for low temperatures, the significant changes of the $-(\frac{\partial M}{\partial T})|_{B}$ from negative to positive values occurs when the magnetic field is applied. The quantum phase transition at $B_{c1}\simeq6.99(5)~T$ and $B_{c2}\simeq14.4(1)~T$ are specified by these sign changes. For better understanding,  $(\partial M/\partial T)=0$ as a function of magnetic field and temperature obtained by mean-field approximation is shown in Fig.~\ref{5}(e). By increasing the temperature, the height of the peaks reduces until all of them are disappearing and no sign of quantum phase transitions can be seen. This indicates that thermal fluctuations become strong enough to take the system to the excited state.

 It should be noted that, as illustrated in Fig.~\ref{5}(b) and (d), the full-diagonalization results shows finite size effects especially in low temperatures and LL regime, but in general the mean-field results are in good agreement with the full-diagonalization and also experimental results\cite{Ruegg08}.


\begin{figure}[t]
\centerline{\psfig{file=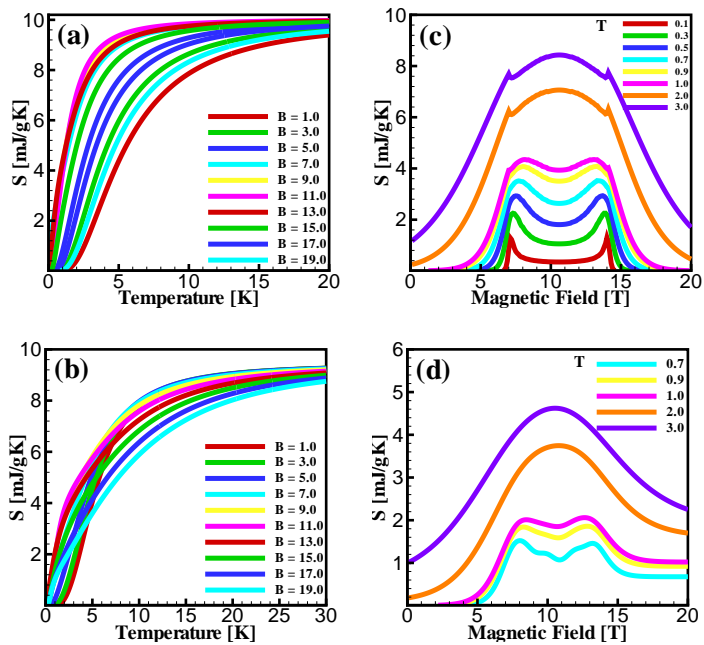,width=3.65in}}
\caption{(color online). Entropy for ladder with
$J_{\perp}=13.0 K$, $J_{\parallel}=3.25 K$ as a function of (a-b)temperature in fixed magnetic field.
(c-d) magnetic field in fixed temperature. (a-c mean-field), (b-d full-diagonalization for $N=16$).}
\label{6}
\end{figure}

~~~~~~~~~~~~~~~~~~~~~~~~~~~~~~~~~~~~~~~~~~~~~~~~~~~~~~~~~~~~~~
~~~~~~~~~~~~~~~~~~~~~~~~~~~~~~~~~~~~~~~~~~~~~~~~~~~~~~~~~~~~~~~~~~~~~~~~~~~~~~~~~~~~~~~~~~~~~~
~~~~~~~~~~~~~~~~~~~~~~~~~~~~~~~~~~~~~~~~~~~~~~
\subsection{Entropy}

In this part, the behavior of the entropy as a function of both temperature and magnetic field is studied. Fig.~\ref{6}(a) and (b) shows the temperature dependence of the entropy obtained by mean-field approach and full-diagonalization method, respectively. At zero temperature, the entropy is zero, which is consistent due to the second law of thermodynamics. When the temperature is high enough, the entropy approaches a constant value, in a fixed magnetic field. Also because of the presence of the gap in $B<B_{c1}$ and $B>B_{c2}$, the entropy is expected to be exponentially activated as a function of temperature.

As it shows in Fig.~\ref{6}(c) and (d), for very low temperatures we can see several interesting results. The maximums of the field dependence of the entropy occur in the critical fields and also the entropy is symmetric about $B=\frac{B_{c1}+B_{c2}}{2}=10.7 T$ just like the specific heat. In very low temperatures, $T<1$, the accumulation of entropy near the quantum critical points is obvious. It implies  that the system is maximally undecided which ground state to choose. As the temperature is enhanced, all the characteristic behaviors have been vanished.

\section{conclusion}\label{sec-5 }

In this paper we have focused on the low-temperature physics of the isotropic spin $S=1/2$ two-leg ladder compound $(C_{5}H_{12}N)_{2} CuBr_{4}$. Using the analytical and numerical methods we have calculated entropy and the specific heat of the system. It is shown that the temperature-dependence of specific heat shows various behaviors in different regions of the magnetic field. The field- dependence of the specific heat shows interesting behavior in the low temperature region. It is almost symmetric about the average of critical points in complete agreement with experimental results.
On the other hand, an external magnetic field induces large relative changes in the entropy of quantum spin systems at finite temperature. This leads to magnetocaloric effect, i.e. a change in temperature during an adiabatic demagnetization process. We computed the entropy of antiferromagnetic spin $S=1/2$ two-leg ladder system in an external magnetic field. Our results show, during an adiabatic demagnetization process temperature drops in the vicinity of a field induced zero-temperature quantum phase transitions.

~~~~~~~~~~~~~~~~~~~~~~~~~~~~~~~~~~~~~~~~~~~~~~~~~~~~~~~~~~~~~~~~~~~~~~~~~~~~~~~~~~~~~~~~~
~~~~~~~~~~~~~~~~~~~~~~~~~~~~~~~~~~~~~~~~~~~~~~~~~~~~~~~~~~~~~~~~~~~~~~~~~~~~~~~~~~~~~~~~~~~~~~~~~~~~~~~~~
~~~~~~~~~~~~~~~~~~~~~~~~~~~~~~~~~~~~~~~~~~~~~~~~~~~~~~~~~~~~~~~~~~~~~~~~~~~~~~
~~~~~~~~~~~~~~~~~~~~~~~~~~~~~~~~~~~~~~~~~~~~~~~~~~~~~~~~~~~~~~~~~~~~~~~~~~~~~~~~~~
~~~~~~~~~~~~~~~~~~~~~~~~~~~~~~~~~~~~~~~~~~~~~~~~~~~~~~~~~~~~~~~~~~~~~~~~~~~~

\section{acknowledgments}
It is our pleasure to thank T. Vekua, H. Hadipour and J. Vahedi for very useful comments and interesting discussions.



\vspace{0.3cm}

\end{document}